\documentclass[superscriptaddress,oneside,twocolumn,showpacs,pre]{revtex4}
\usepackage{graphicx} 
\usepackage{amsmath}  
\begin{document}

\title{Droplet motion driven by surface freezing or melting: A mesoscopic hydrodynamic approach}
\author{Arik Yochelis}
\email{yochelis@technion.ac.il} \affiliation{Department of Chemical Engineering, Technion -- Israel
Institute of Technology, 32000 Haifa, Israel}
\author{Len M. Pismen}
\email{pismen@technion.ac.il} \affiliation{Department of Chemical Engineering, Technion -- Israel
Institute of Technology, 32000 Haifa, Israel} \affiliation{Minerva Center for Nonlinear Physics of
Complex Systems, Technion -- Israel Institute of Technology, 32000 Haifa, Israel}

\received{\today}
\begin{abstract}
A fluid droplet may exhibit self-propelled motion by modifying the wetting properties of the
substrate. We propose a novel model for droplet propagation upon a terraced landscape of ordered
layers formed as a result of surface freezing driven by the contact angle dependence on the terrace
thickness. Simultaneous melting or freezing of the terrace edge results in a joint droplet-terrace
motion. The model is tested numerically and compared to experimental observations on long-chain
alkane system in the vicinity of the surface melting point.
\end{abstract}
 \maketitle

Motion of mesoscopic liquid droplets is a challenging problem both in view of numerous
technological application in surface treatment, microfluidics, etc., and fundamental questions
arising on the borderline between macroscopic and molecular physics. Different scenarios of droplet
motion are determined by liquid-substrate interactions, and may incorporate surface phase
transitions and chemical reactions, as well as more subtle modification of physical properties in
interfacial regions. One can distinguish between three classes of behavior:  \emph{passive},
\emph{interacting}, and \emph{active}. A \emph{passive} droplet gains mobility due to
imposed forces, e.g. temperature gradients~\cite{temp} or substrate heterogeneity~\cite{chem}.
Motion of \emph{interacting} droplets is mediated by fluxes through a thin precursor
layer~\cite{phf04,marmur}. Finally, \emph{active} droplets may propel themselves by modifying the
substrate either through surfactant deposition at the three-phase contact line~\cite{surf} or
through chemical reaction proceeding directly on the substrate at the foot of the
droplet~\cite{sub,TJB:04}.

\begin{figure}[tp]
  \includegraphics[width=3.3in]{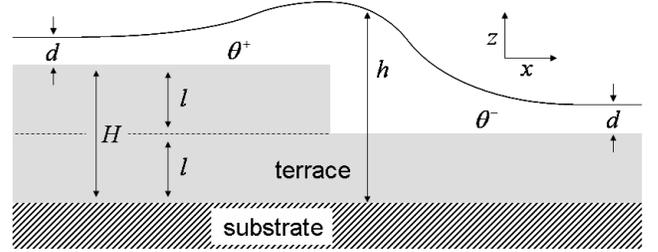}
  \caption{A scheme of a droplet on a terraced smectic layer. $l$ and $d$
denote, respectively, the molecular dimensions along and across the long molecular axis; $H=Nl$ is
the terrace height, $N$ is an integer that denotes number of layers, $h$ is the bulk droplet
height, and $\theta^\pm$ are the contact angles on the upper and lower plateaus.
  }
\label{fig:system}
\end{figure}

A new type of self-propelled motion discovered recently in experiments with long-chain alkanes
($C_nH_{2n+2}$)~\cite{Rieg} is associated with surface phase transitions creating a terraced
immobilized layer between the fluid and substrate, as shown schematically in Fig.~\ref{fig:system}.
The system includes (a) a disordered bulk liquid alkane droplet; (b) one or more ordered (smectic
A) alkane layers formed as a result of surface freezing; (c) a molecularly thin disordered
precursor layer. The thickness ratio $ l /d \gg 1$ of the smectic and precursor layers is
determined by the aspect ratio of the alkane molecule. The plateau height is $H=N l$, where $N \ge
1$ is an integer. A similar situation may arise in layered adsorption, leading to the formation of
ordered immobilized molecular layers also in the case of the aspect ratio $ l /d \sim {\cal O}(1)$.

Due to a difference in molecular interaction strengths between the bulk fluid and the smectic and
the substrate, the contact angle of the bulk droplet depends on the number of smectic layers, and
therefore the droplet is expected to move when placed on a terrace as in Fig.~\ref{fig:system}.
Moreover, as temperature is varied, the terrace surface freezing process may proceed in two ways,
depending on whether it is limited by material supply or removal of latent heat. The first
mechanism involves slow spreading, with the smectic layer growing sidewise, being augmented by
fluid molecules migrating from the bulk droplet through the precursor~\cite{LSR:05}. The second
mechanism is fast, and involves terrace growth synchronous with the droplet motion~\cite{Rieg}.
Melting, being unconstrained by material supply, proceeds by the second mechanism in reverse. In
this Communication, we suggest a model of self propelled droplet motion accompanied by surface
freezing or melting on terraced landscape.

We adopt the lubrication approximation, which accounts for different scalings in the vertical and
the horizontal directions \cite{ODB:97}. The approximation is applicable in a liquid film with a
large aspect ratio, when the interface is weakly inclined and curved. The scaling is consistent if
one assumes  $\partial_z \sim{\cal O}{(1)}$, $\nabla \sim{\cal O}(\epsilon) \ll 1$, $\partial_t
\sim {\cal O}{(\epsilon^2)}$, where $\nabla$ is the two-dimensional gradient in the horizontal
plane. This scaling also implies a small contact angle, $\theta \sim{\cal O}{(1)}$ and results in a
different order of magnitude of the vertical and horizontal velocities, $v_z \sim {\cal
O}{(\epsilon^2)}$, $v_x \sim v_y \sim {\cal O}{(\epsilon)}$. As a consequence, the pressure or,
more generally, a driving potential $W$ is constant to ${\cal O}{(\epsilon^2)}$ across the layer in
$z$ direction. The governing equation for the droplet height $h$ following the  mass conservation
condition reads:
\begin{equation}\label{eq:CH}
    \partial_t h=-\nabla \cdot \mathbf{j}\,,~~~\mathbf{j}= \eta^{-1}k(h)\nabla W\,,
    ~~~W= \Pi-\sigma  \epsilon^2 \nabla^2 h\,,
\end{equation}
where $\mathbf{j}$ is the mass flux, $\eta$ is the dynamic viscosity, $k(h)$ is the mobility
coefficient, $\sigma $ is the surface tension, and $\Pi$ is the disjoining potential due to
interaction with the solid support (including both substrate and smectic layers).

Computation of $\Pi$ is the key component of the model. We assume that all interactions are of
the van der Waals type with the hard core potential $V(r)=\infty$ at $r<d$, $V(r)=-A_jr^{-6}$ at
$r>d$, and differ by interaction constants $A_j$ only.  Since the motion is, on the one hand,
caused by the difference in contact angles on the two sides and, on the other side, is driven by
excess free energy of either freezing or melting, the difference between the liquid-terrace ($A_t$)
and liquid-substrate ($A_s$) interaction constants should change sign when the temperature passes
the surface melting point. Equilibrium contact angles can be expressed through the interaction
constants by integrating Eq.~(\ref{eq:CH}) for an infinite bulk fluid in equilibrium as explained
below.

For a fluid on top of a flat homogeneous plateau, $z>H$ (see Fig.~\ref{fig:system}), the free
energy per unit area can be written, in a local density functional approximation~\cite{Pis:01}, as
\begin{eqnarray}
\gamma &=& \int_H^\infty n (z)\, \left \{ f(n )- \int_{-\infty}^H
Q(z-\zeta)n (z){\rm d} \zeta+ \right.  \nonumber \\ \label{eq:free_enrg} &+&\frac{1}{2}
\int_H^\infty Q(z-\zeta) [n (\zeta)-n  (z)] \, {\rm d} \zeta + \\
\nonumber  &+& \left. \alpha_t n_t \int_{0}^H Q(z-\zeta) {\rm d}
\zeta+\alpha_sn_s \int_{-\infty}^0 Q(z-\zeta) {\rm d} \zeta \right\} {\rm d} z \,.
\end{eqnarray}
Here $n (z)$, $n_t $, $n_s $ are the fluid, terrace, substrate particle densities
and $f(n )$ is free energy per particle of a homogeneous fluid. The first term in the
integrand is the free energy per particle in the homogeneous fluid; the second term compensates lost
fluid-fluid interactions in the domain $z<H$ which are included in $f(n)$; the third term
accounts for the inhomogeneous part of fluid-fluid interactions; the last two terms represent the
fluid-terrace and fluid-substrate interactions. All interactions are described by the same hard
core interaction potential differing only by interaction strength, $A$ (fluid-fluid),
$A_t=\alpha_tA$ (fluid-terrace) and $A_s=\alpha_sA$ (fluid-substrate). The interaction kernel
$Q(\zeta)$ lumping intermolecular interaction between the layers $z=\mathrm{constant}$
\cite{Pis:01} is expressed then as $Q(\zeta)=\frac{1}{2}\pi A \zeta^{-4}$ at $\zeta>d$.

Since the precursor layer is of molecular thickness, the chemical potential shift is computed
differently in the bulk and precursor regions; this is unlike other self-propelled active drop
models~\cite{TJB:04} where a macroscopic precursor layer was presumed. In the \textit{bulk} region
$z>H+d$ the chemical potential shift $\mu(h)-\mu_0$ from the
equilibrium value in the bulk fluid, $\mu=\mu_0$, depends on the fluid thickness $h$ and
coincides with the disjoining potential, $\Pi(h)= \partial_h \gamma$~\cite{der}.
Neglecting the vapor density, as well as density variation in a molecularly-thin interfacial layer,
we can apply the sharp interface approximation~\cite{DiNa:91}, assuming the fluid density to be
constant, $n =n_0  $ at $H+d<z<h$, where $n_0  $ is the equilibrium fluid particle
density at $\mu=\mu_0$, $n =0$ at $z>h$. Defining $\widehat \gamma(h)$ by
Eq.~(\ref{eq:free_enrg}) with the upper integration limit over $z$ replaced by $h$ and the
homogeneous part excluded, we compute
\begin{equation}
\Pi\left(h\right) =\frac{\partial \widehat \gamma}{\partial h}
 =-\frac{\pi A n_0^2}{6}\left[
\frac{\chi}{(h-H)^3}+\frac{\chi_\alpha}{h^3}\right],
\end{equation}
where $\chi=\alpha_t n_t/n_0 -1$ and $\chi_\alpha=(\alpha_s n_s-\alpha_t n_t)/n_0$ are
dimensionless Hamaker constants for fluid-terrace and terrace-substrate interfaces.

The \textit{precursor} film is assumed to be of a \textit{constant} molecular
thickness $d$, but the liquid density is allowed to vary there, and is determined by minimizing
the grand ensemble thermodynamic potential $\mathcal{F} ={\gamma} - \mu \int n \, {\rm d}z$.
The disjoining potential is identified here with the shift of chemical potential per unit volume
$\Pi\left(n \right)=n[\mu(n)-\mu_0]$ relative to the equilibrium value $\mu_0$ as a function of the
local value of $n $ (shifted from its bulk equilibrium value under the action of the terrace and substrate).
It is determined by the Euler--Lagrange equation derived from the integrand of~(\ref{eq:free_enrg}) for $z=H+d$:
\begin{equation}
{\Pi}=n\frac{{\rm d}(n f)}{{\rm d}n}-n\mu_0 -\frac{\pi An}{6}\left[\frac{n_0(\chi+1) -n}{d^3}
+\frac{\chi_\alpha n_0}{(H+d)^3} \right].
\end{equation}

The mobility coefficient, $k(h)$, is also computed separately in the \emph{bulk} and
\emph{precursor} regions and matched at the precursor thickness. In the bulk region, Stokes flow
with a kinematic slip condition~\cite{PR:01} is assumed, while in the precursor domain the mass
transport is presumed to be purely diffusional. This yields the mobility coefficient~\cite{PR:01}
\begin{equation}\label{eq:mobility}
k(h) = \left\{ {\begin{array}{ll}
   {\begin{array}{lll}
   {\lambda^2\left(h-H\right)+\frac{1}{3}\left[h-\left(H+d\right) \right]^3}& {{\rm{at}}} & {h > H+d;} \\
   \end{array}}  \\
   {\begin{array}{lll}
   {\lambda^2}d & {{\rm{at}}} & {h \le H+d} \\
   \end{array}},  \\
\end{array}}
\right.
\end{equation}
where $\lambda=\sqrt{D\eta/n_0  k_BT} \sim \mathcal{O}(d)$ is the effective slip length;
$D$ is surface diffusivity, $k_B$ is Boltzmann constant, and $T$ is temperature.

The motion of a droplet placed on terraced landscape as in Fig.~\ref{fig:system} can be attributed
to a difference in equilibrium contact angles at the upper ($H^+$) and lower ($H^-$) terraces. The
rescaled angles can be calculated for $\chi <0$,  $\vert \chi \vert \ll 1$ by integrating the
static equation $W=0$ \cite{Pis:01}, which reduces to $\sigma \epsilon^2 h_{xx} = \Pi$. In the
limit $h \to \infty$ we obtain
\begin{equation}\label{eq:cont_angl}
    \theta^\pm = \sqrt {\frac{2}{\epsilon^{2}\sigma} \int^\infty_{h_0} \Pi~ {\rm d}h}
    =\sqrt{\frac{\pi A n^2 \vert \chi \vert}{\epsilon^{2}6\sigma d^2}}
    \sqrt{1- \frac{\chi_\alpha/ \vert \chi \vert}{(1+H^\pm/d)^2}}\,,
\end{equation}
where $h_0\approx H^\pm+d$. The direction of the droplet motion is determined solely by the
effective terrace--substrate interaction, i.e. by the sign of $\chi_\alpha$: the droplet either
ascends for $\chi_{\alpha}>0$ or descends for $\chi_{\alpha}<0$ until equilibrium is reached. The
equilibrium condition $\theta^+=\theta^-$ is satisfied either by $H^+=H^-$ or $\chi_\alpha=0$. The
formal small parameter of the lubrication approximation can be defined by setting $\theta=1$ for
$\chi_\alpha=0$, which yields $\epsilon \sim \sqrt{|\chi | An^2/(\sigma d^2)}$. Since $\sigma \sim
n^2 A/d^{2}$, a good estimate is $\epsilon \sim \sqrt{\vert \chi \vert}$.

A decrement of contact angles should be preserved to maintain droplet propagation. This is possible
when the terrace edge is also allowed to move. The terrace motion due to surface freezing or
melting was observed in the experiment ~\cite{Rieg} when ambient temperature $T$ was varied in the
vicinity of the surface freezing point $T_m$. When the terrace is at the foot of a liquid droplet,
the melting or freezing rate is limited by the heat flux $q$ required to supply or remove the
latent heat $L$, so that  $L \rho v = q \approx {\cal K}(T-T_m)/(h-H^+)$, where $\cal K$ is thermal
conductivity and $\rho$ is density (assumed to be equal for both liquid and the frozen terrace
layer). The approximate expression for the heat flux (directed almost normally to the substrate or
terrace) corresponds to the lubrication approximation. The form of this relation defining the edge
position $x$ is
\begin{equation}\label{eq:melt_vel}
    v=\frac{{\rm d}x}{{\rm d}t}= \frac{{\cal K}(T-T_m)}{L \rho \left(h-H^+ \right)}.
\end{equation}

To reproduce joint droplet-terrace dynamics observed in~\cite{Rieg}, we have carried out
dimensionless 1D numerical computations of Eqs.~(\ref{eq:CH}), (\ref{eq:melt_vel}). The new
dimensionless variable forms are: $\widehat h=h/d$, $\xi=x\epsilon/d$, $\tau=t\epsilon^4 \sigma/(d
\eta)$ and $\widehat \Pi=\Pi d/(\epsilon^2 \sigma)$. The particle densities are scaled by $1/b$,
where $b=2\pi d^3/3$ is the excluded volume so that the respective dimensionless equations are:
\begin{subequations}\label{eq:non_dimen}
\begin{eqnarray}
  \label{eq:CH_nondimen}
  \partial_\tau \widehat h=-\partial_\xi k(\widehat h)\partial_\xi \left(\partial_{\xi \xi} \widehat h- \widehat \Pi
  \right)\,,\\
  \widehat v=\frac{{\rm d}\xi}{{\rm d}\tau}=\frac{\Delta}{\widehat h-H^+},
  \label{eq:melt_vel_nondimen}
\end{eqnarray}
where
\begin{equation}\label{eq:dis_potn_nondimen}
{\widehat \Pi}  = \left\{ {\begin{array}{lrr}
   {\begin{array}{lll}
   { - \dfrac{{\beta \widehat n^2_0}}{4}\left[ \dfrac{\chi}{ ({\widehat h - H})^3} + \dfrac{\chi_\alpha }{\widehat h^3} \right]} & \quad  \rm{at} & {\widehat h > H+1;}  \\
   \end{array}}  \\
   {\begin{array}{lll}
   -\dfrac{\beta \widehat n}{4}\left[{\widehat n_0\left(\chi+1 \right)-\widehat n } + \dfrac{{\widehat n_0 \chi_\alpha }}{{\left( {1 + H} \right)^3 }} \right]-\widehat n \widehat \mu_0+& &  \\
   \end{array}}  \\
   {\begin{array}{rrr}
   +\dfrac{\widehat n}{1-\widehat n } - \widehat n\ln\left(\dfrac{1}{\widehat n }- 1 \right)-2 \beta \widehat
   n^2
    & {\rm{at}} & {\widehat h \le H+1}\,,
   \end{array}}
\end{array}}
\right.
\end{equation}
\begin{equation}\label{eq:Delta}
    \Delta=\frac{{\cal K}(T-T_m)}{L \rho d}\frac{d \eta \epsilon}{\epsilon^4 \sigma d}\sim \frac{{\cal K}(T-T_m) \eta}{L \rho d \sigma \vert \chi
    \vert^3},
\end{equation}
\end{subequations}
$\beta=A/(k_BTd^6)$ and $ \widehat \mu_0=\mu_0 b/(k_BT)$. The same notation is retained for the
dimensionless variables, and Eq.~(\ref{eq:mobility}) remains without change, except replacing $d
\to 1$. The density in the precursor domain, $\widehat n $ is transformed to effective height as
$\widehat n =\widehat n_0 (\widehat h-H)$. We adopted the explicit spectral method and by doubling
the grid size impose reflecting boundary conditions. The initial state in each computation includes
a droplet with its maximum placed above the terrace edge and a precursor film of unit thickness
(see Fig.~\ref{fig:system}). The parameter $\Delta$ defining the ratio of characteristic velocities
of the edge to droplet motion is of ${\cal O}(1)$ when the temperature difference is in the range
of ${\cal O}(10^{-3})[^0K]$.
\begin{figure}[tp]
  \includegraphics[width=3.3in]{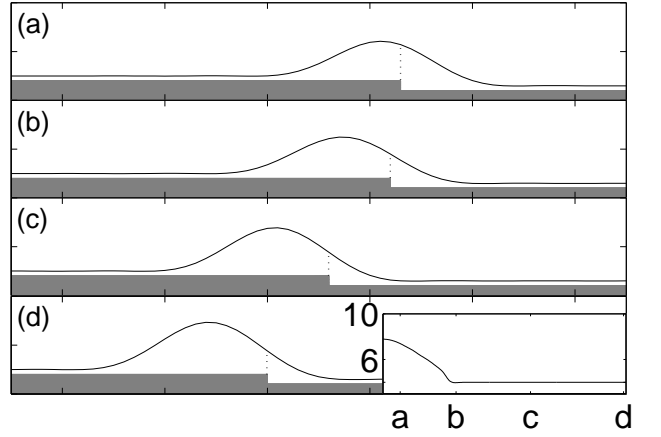}
  \caption{Numerical solution of the fluid and the terrace according to Eq.~(\ref{eq:non_dimen}), showing the melting
  process (a-d) at respective time steps (from top to bottom: 50, 210, 400, and 650).
  The horizontal range is $\xi=[0,120]$ and the vertical range is $\widehat h=[0,20]$.
  The dotted line marks the droplet height above the terrace edge, $\widehat h_c$.
  The inset shows the dependence of $\widehat h_c-H^+$ as a function of time and its relaxation to an equilibrium
  value $\widehat h_c-H^+\simeq 4$.
  Parameters: $\chi=-0.3$, $\beta=15$, $\lambda=\sqrt{3}$, $ l=2$, $H^+=2 l$, $H^-= l$,
  $\chi_\alpha=-10$ and $\Delta=-4$.}
  \label{fig:dropL}
\end{figure}

Synchronous droplet-terrace motion under melting conditions is shown in Fig.~\ref{fig:dropL}. This
joint propagation can be explained in terms of terminal velocity of the terrace. While the terrace
is below the droplet, the droplet velocity is determined by the difference in contact angle values,
according to~(\ref{eq:cont_angl}). On the other hand, the terrace velocity (for a fixed $\Delta$)
depends solely on the thickness of the liquid layer above the terrace edge. At the start, the
droplet moves to the left, while the terrace remains almost stationary because of slow transport
through a thick layer, as shown Fig.~\ref{fig:dropL}(a-b). As the fluid height above the edge
decreases, the terrace gains speed [see Fig.~\ref{fig:dropL}(b-c)], until it reaches an
``equilibrium'' position, such that the point at the droplet interface just above the edge where
the thickness is $\widehat h_c$ moves with the same speed $\widehat v =\Delta/(\widehat h_c-H^+)$.
The stable position should lie near the trailing edge; then, if the terrace moves faster than the
droplet, the liquid layer thickness above it increases and the terrace decelerates. As a result,
the value $\widehat h=\widehat h_c$ remains constant, as seen in Fig.~\ref{fig:dropL}(c-d)] and
more precisely in inset of Fig.~\ref{fig:dropL}(d). This dynamic feedback allows the droplet and
the terrace to synchronize their motion. As implied by~(\ref{eq:cont_angl}), we found that the
synchronous propagation velocity depends on the layers thickness $H^+=Nl$ and will be discussed
elsewhere~\cite{YoPi:05}.
\begin{figure}[tp]
  \includegraphics[width=3.3in]{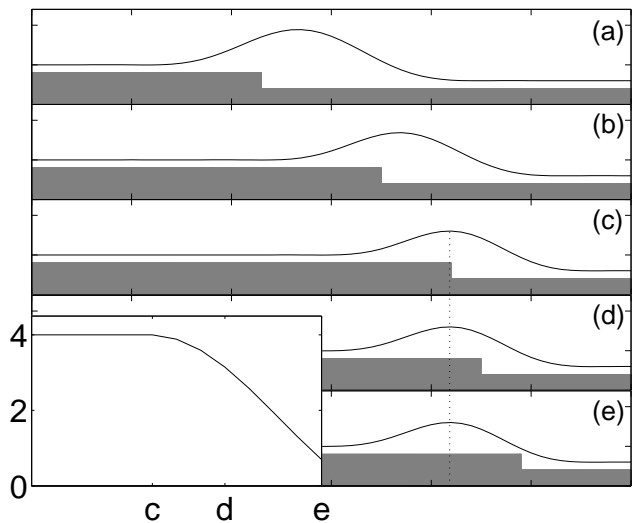}
  \caption{Numerical solution of Eq.~(\ref{eq:non_dimen}), showing the freezing
  process (a-e) at respective time steps (from top to bottom: 0, 410, 581.85, 581.88 and 581.92).
   The horizontal range is $\xi=[0,120]$ and the vertical range is $\widehat h=[0,12]$.
  The dotted line in (c--e) marks the droplet position according to its maximal height.
  The inset shows the dependence of $\widehat h_c-H^+$ on time in the vicinity of the critical droplet volume.
   Parameters: $\chi=-0.3$, $\beta=15$, $\lambda=\sqrt{3}$, $l=2$, $H^+=2 l$, $H^-= l$,
  $\chi_\alpha=1.5$ and $\Delta=4$.}
  \label{fig:dropR}
\end{figure}

In the freezing case, the droplet volume decreases, since the total mass of the system is conserved
and the fluid is solidified. The droplet and the terrace may still jointly propagate as long as the
droplet height is relatively large compared to the precursor thickness, as shown in
Fig~\ref{fig:dropR}(a-c). In a such motion the droplet and the terrace preserve the equilibrium
height $\widehat h=\widehat h_c$ [see Fig.~\ref{fig:dropR}(b-c)]. As the droplet volume decreases
below the equilibrium height $\widehat h_c$, the terrace propagates faster than the droplet and
runs out to its leading edge [see Fig.~\ref{fig:dropR}(d-e)]. Following this, the motion stops,
since further terrace propagation is limited by slow material supply through the precursor, and the
droplet is left in an equilibrium state on the top of a flat smectic layer~\cite{Rieg}. This
behavior is also presented in the inset of Fig.~\ref{fig:dropR}. As the terrace passes the maximum
droplet's height, the critical value $\widehat h_c-H^+$ decreases to unity (i.e. the precursor
thickness). The droplet velocity at the same time drops to zero, while the terrace velocity (dashed
line) jumps abruptly. Similar behavior has been also observed  experimentally~\cite{LSR:05}.

We have proposed a model for self-propelled droplets on top of a terraced landscape driven by
surface freezing or melting. The numerical estimates show the characteristic terrace edge velocity
$v \sim {\cal O}(10^{2})\,[\mu m/sec]$ close to the experimental data~\cite{Rieg} at temperature
variations around the surface melting temperature $\vert T-T_m\vert\sim {\cal O}(10^{-3})[^0K]$ and
$h_c \sim {\cal O}(d)\sim 0.1[nm]$.

We thank Hans Riegler for stimulating discussions and for sharing with us his unpublished material.
This research has been supported by Israel Science Foundation (grant \# 55/02).

\end{document}